# DEFM: Delay-Embedding-based Forecast Machine for Time Series Forecasting by Spatiotemporal Information Transformation

Hao Peng[1], Wei Wang[1], Pei Chen[1*], Rui Liu[1,2*]

[1] School of Mathematics, South China University of Technology, Guangzhou 510640, China.

[2] Pazhou Lab, Guangzhou 510330, China.

* Correspondence: Pei Chen, chenpei@scut.edu.cn; Rui Liu, scliurui@scut.edu.cn

**Abstract**

Making accurate forecasts for a complex system is a challenge in various practical applications. The major difficulty in solving such a problem concerns nonlinear spatiotemporal dynamics with time-varying characteristics. Takens' delay embedding theory provides a way to transform high-dimensional spatial information into temporal information. In this work, by combining delay embedding theory and deep learning techniques, we propose a novel framework, Delay-Embedding-based Forecast Machine (DEFM), to predict the future values of a target variable in a self-supervised and multistep-ahead manner based on high-dimensional observations. With a three-module spatiotemporal architecture, the DEFM leverages deep neural networks to effectively extract both the spatially and temporally associated information from the observed time series even with time-varying parameters or additive noise. The DEFM can accurately predict future information by transforming spatiotemporal information to the delay embeddings of a target variable. The efficacy and precision of the DEFM are substantiated through applications in three spatiotemporally chaotic systems: a 90-dimensional (90D) coupled Lorenz system, the Lorenz 96 system, and the Kuramoto-Sivashinsky (KS) equation with inhomogeneity. Additionally, the performance of the DEFM is evaluated on six real-world datasets spanning various fields. Comparative experiments with five prediction methods illustrate the superiority and robustness of the DEFM and show the great potential of the DEFM in temporal information mining and forecasting.

**With the explosive generation of time series datasets from various complex systems, methods have been presented to solve the challenging time series forecasting problem. Although conventional methods have made progress in time series analysis scenarios, the high-dimensional spatial information contained in time series has not yet been fully exploited. Takens' delay embedding theory offers a technique that enables spatial-temporal information (STI) transformation, based on which we develop a novel framework, i.e., the DEFM, for accurately and robustly forecasting the dynamics of a target variable. Specifically, by transforming the spatial information of a high-dimensional time series to the temporal information of a target variable, the unknown future values of the target can be solved in a multistep-ahead manner via a self-supervised learning scheme. Based on the combination of a DEFM-based STI transformation and deep learning, the DEFM can fully leverage the nonlinear spatial and temporal information acquired from a high-dimensional dynamic system. This work can be regarded as a concrete step toward data-driven research on high-dimensional nonlinear systems, and the proposed approach has great potential for use in real-world applications.**

## 1. Introduction

In the era of big data, enormous amounts of time-course data are observed and generated in many fields, such as gene expression data in molecular biology, neural activity data in neuroscience, and atmospheric data in meteorology[1–7]. In these areas, the availability of accurate empirical models for multistep-ahead prediction is highly desirable. However, in view of the complexity and nonlinearity of real-world systems, it is challenging to bridge known information to future dynamics, especially when the known time series contains a massive number of factors (high-dimensional variables). Numerous approaches have been developed for time series analysis, e.g., statistical regression methods, such as the mean average through exponential smoothing[8,9] and the autoregressive integrated moving average (ARIMA) model[10], and machine learning methods, including recurrent neural networks (RNNs)[11–14], long short-term memory (LSTM) networks[15,16], and reservoir computing[17,18]. However, most existing methods generally only rely on the temporal information of time series, which limits their applications in many real-world systems. In contrast with the numerous existing prediction methods based on time association mining, few studies focusing on leveraging high-dimensional data can be found. Making time series forecasts based on high-dimensional data is usually necessary. On the one hand, high-dimensional time series data always contain spatial heterogeneity in their system dynamics, which can significantly improve the performance of the time series forecasting method and reduce the needed length for the input time series[19]. On the other hand, handling high-dimensional data is a common occurrence in many practical scenarios, such as biological experiments[20]. Thus, new strategies are needed to better explore the time series derived from high-dimensional complex systems.

Takens' delay embedding theory[21,22] reveals that a high-dimensional attractor and reconstructed delay attractors with appropriately selected dimensions are topologically conjugated, thus suggesting mappings from the original attractor to the reconstructed attractors. Derived from delay embedding theory, the randomly distributed embedding (RDE)[23,24] approach makes a one-step-ahead prediction for a target variable based on short-term time series through a spatial-temporal information (STI) transformation. Our recently proposed autoreservoir computing framework (ARNN)[25] achieves multistep-ahead forecasting based on a semi-linearized STI transformation; however, the challenge is how to exploit the nonlinear features of the observed high-dimensional variables, which restricts the forecasting performance of the model in terms of robustness and accuracy.

In this study, we propose a novel neural network-based framework, a delay-embedding-based forecasting machine (DEFM), to produce highly accurate multistep-ahead predictions by transforming the spatial information of a high-dimensional time series to the temporal information of a target variable (Fig. 1 (a)). Motivated by the previous linear or semi-linearized STI-based works[23,25], the DEFM is designed to solve the nonlinear STI equation that maps the observed high-dimensional variables to the delayed embeddings of a target variable (variable to be predicted) in a self-supervised manner. Specifically, the DEFM utilizes its spatiotemporal structure, consisting of a spatial module and a temporal module, to exploit the intrinsic dynamics of a complex system and extract both the spatial interactions and temporally associated information among the variables in high-dimensional data (Fig. 1(b)). With such a data-driven architecture, the DEFM takes the high-dimensional variables as inputs and provides the delay embeddings of the target variable as outputs, and it is trained by a "consistently self-constrained scheme", thus directly producing a multistep-ahead prediction for the target variable with only one forward step (Fig. 1 (b)). Simultaneously, the DEFM can be applied to make long-term predictions through an iterative scheme that takes the predicted values as part of the input for the next prediction step. We present DEFM applications across several benchmark systems, underscoring the robustness and effectiveness of the approach. Through real-world applications, our investigation highlights the capacity of the DEFM to serve as a valuable analytical instrument for the examination and prediction of intricate dynamic systems. Notably, our method relies on high-dimensional time series data, emphasizing its versatility and applicability in practical scenarios.

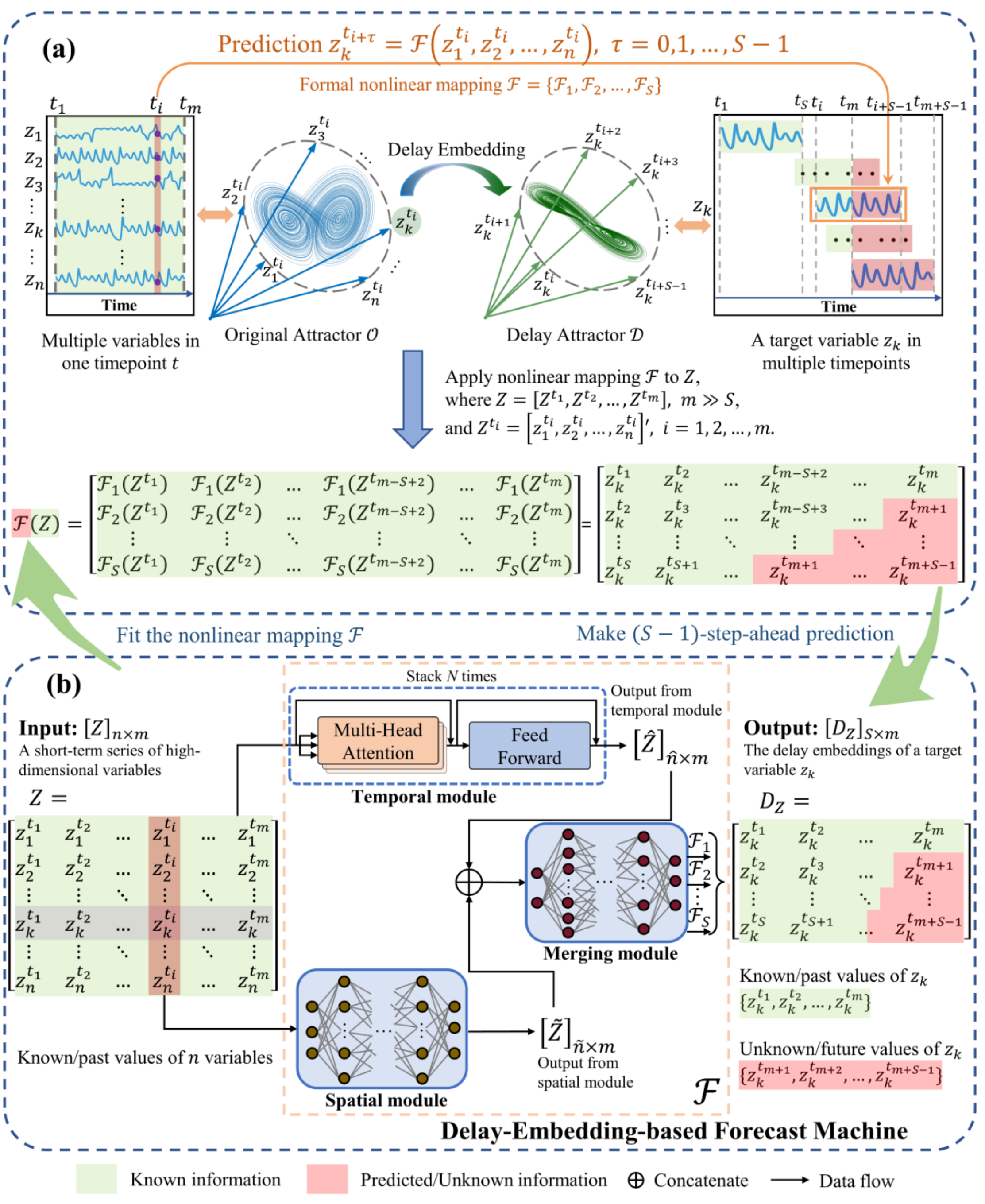

**Fig. 1.** Schematic illustration of the DEFM. **(a)** Through a delay embedding scheme[21,22], a delay attractor $\mathcal{D}$ is reconstructed for a target variable $z_k$ with appropriately reconstructed dimensions, and this attractor is topologically conjugated with the original attractor $\mathcal{O}$. From attractor $\mathcal{O}$ to attractor $\mathcal{D}$, a formal nonlinear function $\mathcal{F}$ maps the known spatial information to the temporal information of the target variable $z_k$, while the future values of $z_k$ are obtained by $z_k^{t_{i+\tau}} = \mathcal{F}\left(z_1^{t_i}, z_2^{t_i}, \dots z_n^{t_i}\right), \tau = 1,2,\dots,S-1$. Therefore, the key to predicting $z_k$ is to fit the nonlinear function $\mathcal{F}$. **(b)** For the given target variable $z_k$, the DEFM framework is designed to solve the DEFM-based STI equation (Eq. (4)) and thus offers a bridge between the high-dimensional data $Z$ and the delay embedding matrix $D_z$. The DEFM fully extracts rich spatial and temporal information from high-dimensional data with temporal and spatial modules and integrates this spatiotemporal information via a merging module. After being trained in a self-supervised way, the DEFM predicts the future information of $z_k$ in a multistep-ahead

manner, i.e., simultaneously providing the unknown values $\{z_k^{t_{m+1}}, z_k^{t_{m+2}}, \dots, z_k^{t_{m+S-1}}\}$ in the output delay embedding matrix $D_z$.

## 2. Methods

### 2.1 Delay embedding theorem for dynamical systems

The dynamics of a general discrete-time dissipative system can be defined as:

$$Z^{t_{i+1}} = \phi(Z^{t_i}), \tag{1}$$

where $\phi: \mathbb{R}^n \to \mathbb{R}^n$ is a nonlinear mapping, and its variables are defined in a $n$-dimensional state space $Z^{t_i} = \left(z_1^{t_i}, z_2^{t_i}, \dots, z_n^{t_i}\right)^T$ at a time point $t = t_i$, where the symbol " $T$ " is the transpose of a vector, and any time interval between two consecutive time points is equal. After a sufficiently long time, all states converge to a compact manifold $\mathcal{V}$. Denoting $b$ as the box-counting dimension for the attractor contained in manifold $\mathcal{V}$, the delay embedding theorem indicates that only the observed long-term data of a single variable are needed to topologically reconstruct the attractor of the original high-dimensional system when certain conditions are satisfied. Takens' embedding theorem[21,22] is formally articulated in Theorem 1 as follows.

**Theorem 1.** If $\mathcal{V}$ is a compact manifold or an attractor with a dimensionality of $b$, for a smooth diffeomorphism $\phi: \mathcal{V} \to \mathcal{V}$ and a smooth function $g: \mathcal{V} \to \mathbb{R}^1$, there is a generic property stating that the mapping $\mathcal{F}_{\phi,g}: \mathcal{V} \to \mathbb{R}^S$ is an embedding when $S > 2b$, that is:

$$\mathcal{F}_{\phi,g}(Z) = [g(Z), g \circ \phi(Z), \dots, g \circ \phi^{S-1}(Z)]^T, \tag{2}$$

where the symbol " $\circ$ " is the function composition operation.

In particular, letting $Z = Z^{t_i}$ and $g \circ \phi^j(Z^{t_i}) = z_k^{t_{i+j}}$, where $z_k^{t_{i+j}} \in \mathbb{R}^1$, $z_k$ is anyone among $n$ variables $\{z_1, z_2, \dots, z_n\}$, and $j = 0,1, \dots, S-1$, the mapping above has the following form with $\mathcal{F} = \mathcal{F}_{\phi,g}$:

$$\mathcal{F}(Z^{t_i}) = \left[z_k^{t_i}, z_k^{t_{i+1}}, \dots, z_k^{t_{i+S-1}}\right]^T. \tag{3}$$

### 2.2 DEFM-based STI transformation

As shown in Fig. 1 (a), the original observed attractor in an $n$-dimensional space is defined as $\mathcal{O}\left(z_1^{t_i}, z_2^{t_i}, \dots, z_n^{t_i}\right)$ at time points $t = t_i\ (i = 1, 2, \dots, m)$, where $\{z_j^{t_1}, z_j^{t_2}, \dots, z_j^{t_m}\}$ is the observed time series of each variable $z_j\ (j = 1, 2, \dots, n)$, $n$ is the number of variables, and $m$ is the number of time points in the observed time series. Therefore, for this dynamical system, an observed $n$-dimensional time series is denoted as a matrix $Z = [Z^{t_1}, Z^{t_2}, \dots, Z^{t_m}]_{n\times m}$, where $Z^{t_i} = \left[z_1^{t_i}, z_2^{t_i}, \dots, z_n^{t_i}\right]^T \in \mathbb{R}^n$ is a high-dimensional/spatial vector containing $n$ observed variables at time point $t = t_i$ $(i = 1, 2, \dots, m)$.

A target variable (variable to be predicted) $z_k$ is any one of the $n$ variables $\{z_1, z_2, \dots, z_n\}$. Based on Takens' delay embedding theory[21,22], a topologically equivalent delay attractor is reconstructed in the form of $\mathcal{D}(z_k^{t_i}, z_k^{t_{i+1}}, \dots, z_k^{t_{i+S-1}})$, where the embedding dimensionality $S$ satisfies $S > 2b > 0$, with $b$ as the box-counting dimensionality of the original attracter $\mathcal{O}$ (see Supplementary Information Section 1.1 for details). In such a way, the higher-dimensional dynamics governed by massive spatial variables $\{z_1, z_2, \dots, z_n\}$ are "folded" into a low-dimensional embedding space enriched with the temporal information $\{z_k^{t_1}, z_k^{t_2}, \dots, z_k^{t_{m+S-1}}\}$ of a variable $z_k$. Thus, a formal nonlinear mapping $\mathcal{F} = \{\mathcal{F}_l\}_{l=1}^{S}$ with sub-mappings $\mathcal{F}_l(Z^{t_i}) = z_k^{t_{i+l-1}}, l = 1,2, \dots, S$ is constructed from the high-dimensional variables $\{z_1, z_2, \dots, z_n\}$ to the temporal delay embeddings of the target variable $z_k$, as shown in Fig. 1 (a) and Eq. (4).

$$\mathcal{F}(Z) = \begin{bmatrix} \mathcal{F}_1(Z^{t_1}) & \mathcal{F}_1(Z^{t_2}) & \cdots & \mathcal{F}_1(Z^{t_m}) \\ \mathcal{F}_2(Z^{t_1}) & \mathcal{F}_2(Z^{t_2}) & \cdots & \mathcal{F}_2(Z^{t_m}) \\ \vdots & \vdots & \ddots & \vdots \\ \mathcal{F}_S(Z^{t_1}) & \mathcal{F}_S(Z^{t_2}) & \cdots & \mathcal{F}_S(Z^{t_m}) \end{bmatrix} = \begin{bmatrix} z_k^{t_1} & z_k^{t_2} & \cdots & z_k^{t_m} \\ z_k^{t_2} & z_k^{t_3} & \cdots & z_k^{t_{m+1}} \\ \vdots & \vdots & \ddots & \vdots \\ z_k^{t_S} & z_k^{t_{S+1}} & \cdots & z_k^{t_{m+(S-1)}} \end{bmatrix} = D_z, \tag{4}$$

where $\mathcal{D}_z$ is an $S \times m$ matrix constructed by a series of delay embeddings. Eq. (4) is also called the DEFM-based STI equation, in which the high-dimensional known/past information includes the time series $\{z_j^{t_1}, z_j^{t_2}, \dots, z_j^{t_m}\}$ of each variable $z_j\ (j =$ $1, 2, \ldots, n)$, while the unknown/future values are $\{z_k^{t_{m+1}}, z_k^{t_{m+2}}, \ldots, z_k^{t_{m+S-1}}\}$ (in the shadow area) for the target variable $z_k$ (Fig. 1 (a)).

Clearly, fitting the formal nonlinear mapping $\mathcal{F}$ is the key to precisely predicting the future values of the target variable $z_k$ from the known information $Z$.

### 2.3 The architecture of the DEFM

First, based on Eq. (4), $Z = [Z^{t_1}, Z^{t_2}, \ldots, Z^{t_m}]_{n \times m}$ is the input matrix, $Z^{t_i} = \left[z_1^{t_i}, z_2^{t_i}, \ldots, z_n^{t_i}\right]^T \in \mathbb{R}^n$ is a vector containing $n$ observed variables at time point $t = t_i$ $(i = 1, 2, \ldots, m)$. The delay embedding matrix $[D_z]_{S \times m}$ is the output containing both the known series $\{z_k^{t_1}, z_k^{t_2}, \ldots, z_k^{t_m}\}$ and the future series $\{z_k^{t_{m+1}}, z_k^{t_{m+2}}, \ldots, z_k^{t_{m+L-1}}\}$ of the target variable $z_k$.

To explore the intertwined information in observed high-dimensional time series data, a spatiotemporal structure is deployed for the DEFM framework, as shown in Fig. 1 (b). The DEFM is composed of three modules, including temporal, spatial, and merging modules. The temporal module is assembled by a series of so-called self-attention[26] layers, which captures the underlying sequentially and temporally associated information among the input samples across all $m$ observed time points (Fig. 1 (b)). The spatial module for extracting the interactions among $n$ variables in the high-dimensional data and a merging module for processing the spatiotemporal information and outputting the delay time series are both implemented by two separate fully connected neural networks. Specifically, given a matrix $Z \in \mathbb{R}^{n \times m}$ consisting of the time series data of the high-dimensional variables $\{z_1, z_2, \ldots, z_n\}$ as the input, two data flow branches are utilized; i.e., each column $Z^{t_i} = \left[z_1^{t_i}, z_2^{t_i}, \ldots, z_n^{t_i}\right]^T$ of $Z$ (a spatial vector at a time point $t = t_i$) is input into the spatial module, while the rows (the time series across all $m$ time points $\{t_1, t_2, \ldots, t_m\}$) are fed into the temporal module (Fig. 1 (b)). Then, the outputs of the temporal branch $\hat{Z} \in \mathbb{R}^{\hat{n} \times m}$ and that of spatial branch $\tilde{Z} \in \mathbb{R}^{\tilde{n} \times m}$ are aggregated into one feature matrix through a concatenation operation and inputted into the merging module (Fig. 1 (b)).

The temporal module of the DEFM is assembled by $N$ self-attention[26] layers, which are applied to determine the temporal interactions among variables[26–28]. In particular, each self-attention layer contains a multi-head attention layer and a Feed-Forward layer. First, the high-dimensional $Z^{t_i} \in \mathbb{R}^n$ $(i = 1,2,\ldots,m)$ is projected to a query vector $q^{t_i} \in \mathbb{R}^{d_a}$, a key vector $k^{t_i} \in \mathbb{R}^{d_a}$ and a value vector $v^{t_i} \in \mathbb{R}^{d_a}$, where $d_a$ is the dimensionality of the attention model. Note that $Q = [q^{t_1}, q^{t_2}, \ldots, q^{t_m}]^T, K = [k^{t_1}, k^{t_2}, \ldots, k^{t_m}]^T, V = [v^{t_1}, v^{t_2}, \ldots, v^{t_m}]^T$. In each head subspace, the queries, keys, and values are produced by linear projections with $d_q, d_k$ and $d_v$ dimensions. The final multi-head attention values are determined by the concatenated parallel attention values derived from the $h$ heads in the following form:

$$\mathrm{MultiHead}(Q, K, V) = \mathrm{Concat}(\mathrm{head}_1, \ldots, \mathrm{head}_h)W^o, \tag{5}$$

with

$$\begin{aligned} \mathrm{head}_j &= \mathrm{Attention}\left(QW_j^Q, KW_j^K, VW_j^V\right) \\ &= \mathrm{softmax}\left(\frac{\left(QW_j^Q\right)\left(KW_j^K\right)^T}{\sqrt{d_k}}\right)\left(VW_j^V\right), \end{aligned} \tag{6}$$

where $W^o \in \mathbb{R}^{hd_v \times d_a}, W_j^Q \in \mathbb{R}^{d_a \times d_q}, W_j^K \in \mathbb{R}^{d_a \times d_k}$ and $W_j^V \in \mathbb{R}^{d_a \times d_v}$ are the weight matrices. The Feed-Forward layer is usually a fully connected network, which is adopted to separately process each feature. Thus, each column $\hat{Z}^{t_i} \in \mathbb{R}^{\hat{n}}$ in the output of the temporal module $\hat{Z} \in \mathbb{R}^{\hat{n} \times m}$ (Fig. 1 (b)), which is computed by $N$ stacked self-attention layers, contains the temporal features at time $t = t_i, i = 1,2,\ldots,m$, suggesting that the temporal module with stacked self-attention layers explores the sequentially and temporally associated information among the input samples across all $m$ time points.

The spatial module is a neural network that takes each column $Z^{t_i} = \left[z_1^{t_i}, z_2^{t_i}, \ldots, z_n^{t_i}\right]^T$ of $Z$ (a spatial vector at a time point $t = t_i$) as input neurons and outputs the spatial features $\tilde{Z}^{t_i}$ at time $t = t_i$. Aggregated by concatenating the temporal features $\hat{Z}$ and spatial features $\tilde{Z}$, each column of the merged feature matrix includes the spatiotemporal information at time $t = t_i$. The spatial module and merging module are implemented by fully connected neural networks.

Finally, the delay embedding matrix $D_z$ of the target variable $z_k$ is the output of the merging module. To integrate $S$ sub-mapping functions, an $S$-node dense layer is adopted at the end of the merging module. Utilizing the architecture with the above three modules, the DEFM can efficiently transform the spatial information derived from a high-dimensional time series into temporal information and provide the future values $\{z_k^{t_{m+1}}, z_k^{t_{m+2}}, \ldots, z_k^{t_{m+S-1}}\}$ of the target variable.

### 2.4 Self-supervised training process of the DEFM

Based on the DEFM architecture, the training samples are paired vectors, i.e., $\left[z_1^{t_i}, z_2^{t_i}, \ldots, z_n^{t_i}\right]^T$ and $\left[z_k^{t_i}, z_k^{t_{i+1}}, \ldots, z_k^{t_{i+S-1}}\right]^T$ with $i = 1, 2, \ldots, m$. Remembering that the nonlinear function $\mathcal{F}$ is composed of $S$ intertwined sub-mappings $\{\mathcal{F}_1, \mathcal{F}_2, \ldots, \mathcal{F}_S\}$, the DEFM is trained by a "consistently self-constrained scheme" to fit the sub-mappings simultaneously, thus maintaining the integrity of $\mathcal{F}$ and preserving temporal consistency. Specifically, due to the nature of the delay embeddings in the output $D_z$ (as shown in Eq. (4)), we have a total of $m + S - 3$ temporally self-constrained conditions:

$$\mathcal{F}_{j-1}(Z^{t_{i+1}}) = \mathcal{F}_j(Z^{t_i}), \tag{7}$$

where $j \in \{2, 3, \ldots, S\}$ and $Z^{t_i} = \left[z_1^{t_i}, z_2^{t_i}, \ldots, z_n^{t_i}\right]^T$ is a spatial sample at time point $t = t_i$ $(i = 1, 2, \ldots, m)$. Among these conditions, $m - 1$ conditions are imposed for the known states, and $S - 2$ conditions are applied for the future values. Clearly, these cross-sample conditions in Eq. (7) constrain the training process in terms of the temporal sample sequence. For the target variable $z_k$, the estimated values of its delay embeddings $\widehat{D}_z$ are obtained from the known data matrix $Z$:

$$\widehat{D}_z = \begin{bmatrix} \left(\hat{z}_k^{t_1}\right)_1 & \left(\hat{z}_k^{t_2}\right)_1 & \cdots & \left(\hat{z}_k^{t_m}\right)_1 \\ \left(\hat{z}_k^{t_2}\right)_2 & \left(\hat{z}_k^{t_3}\right)_2 & \cdots & \left(\hat{z}_k^{t_{m+1}}\right)_2 \\ \vdots & \vdots & \ddots & \vdots \\ \left(\hat{z}_k^{t_S}\right)_S & \left(\hat{z}_k^{t_{S+1}}\right)_S & \cdots & \left(\hat{z}_k^{t_{m+S-1}}\right)_S \end{bmatrix}, \tag{8}$$

where $\left(\hat{z}_k^{t_i}\right)_j$ $(i = 1, 2, \ldots, m + S - 1;\ j = 1, 2, \ldots, S)$ is generated from the output of the $j^{\text{th}}$ sub-mapping function $\mathcal{F}_j$. From the matrix $\widehat{D}_z$ estimated above, the unknown future values of the target variable $\{z_k^{t_{m+1}}, z_k^{t_{m+2}}, \dots, z_k^{t_{m+S-1}}\}$ are provided by Eq. (9):

$$z_k^{t_{m+i}} = \frac{1}{S-i} \sum_{j=i+1}^{S} \left(\hat{z}_k^{t_{m+i}}\right)_j . \tag{9}$$

Through an autoperception procedure, the training and optimization steps of the DEFM are accomplished through a process that minimizes a loss function $\mathcal{L} = \mathcal{L}_{DS} + \mathcal{L}_{FC}$, where $\mathcal{L}_{DS}$ is a determined-state loss derived from the observed/known states $\{z_k^{t_1}, z_k^{t_2}, \dots, z_k^{t_m}\}$ of $z_k$, and $\mathcal{L}_{FC}$ is a future-consistency loss in terms of the future/unknown series $\{z_k^{t_{m+1}}, z_k^{t_{m+2}}, \dots, z_k^{t_{m+S-1}}\}$ of $z_k$. Specifically, the determined-state loss $\mathcal{L}_{DS}$ is defined as:

$$\mathcal{L}_{DS} = \frac{1}{2mS - S^2 + S} \sum_{j=1}^{S} \sum_{i=j}^{m} \left( \left(\hat{z}_k^{t_i}\right)_j - z_k^{t_i} \right)^2 , \tag{10}$$

where $\left(\hat{z}_k^{t_i}\right)_j$ is the estimation of $\mathcal{F}_j(Z^{t_i})$ with $Z^{t_i} = \left[z_1^{t_i}, z_2^{t_i}, \dots, z_n^{t_i}\right]^T$, and $z_k^{t_i}$ $(i = 1, 2, \dots, m)$ is the known value of $z_k$. The loss $\mathcal{L}_{DS}$ is constructed from the differences between the estimations $\left(\hat{z}_k^{t_i}\right)_j$ and the observed values (ground truths) $z_k^{t_i}$ for all past time points $t = t_i$ $(i = 1, 2, \dots, m)$. The future-consistency loss $\mathcal{L}_{FC}$ is defined as:

$$\mathcal{L}_{FC} = \frac{1}{S(S-1)} \sum_{j=2}^{S} \sum_{i=m+1}^{m+j-1} \left( \left(\hat{z}_k^{t_i}\right)_j - \operatorname{mean}\left(\left(\hat{z}_k^{t_i}\right)\right) \right)^2 , \tag{11}$$

where $\operatorname{mean}\left(\left(\hat{z}_k^{t_i}\right)\right)$ denotes the average of all estimated future values of $z_k$ that correspond to the same future time point $t_i$ $(i = m+1, m+2, \dots, m+S-1)$. Clearly, $\mathcal{L}_{FC}$ is constructed from the temporally self-constrained conditions in Eq. (7). An intuitive understanding of the future-consistency loss is that minimizing $\mathcal{L}_{FC}$ ensures that the outputs obtained from different sub-mappings but corresponding to the same future time point $t = t_i$ are identical, which preserves the temporal consistency of the outputs in the lower-right corner of the delay embedding matrix $D_z$ during the training procedure.

Based on the integration of the above two losses, the DEFM is trained in a self-supervised manner, i.e., without any requirement for future labels of the target variable. The cooperation between the future-consistency loss $\mathcal{L}_{FC}$ and the determined-state loss $\mathcal{L}_{DS}$ helps to solve the DEFM-based STI equation (Eq. (4)). The detailed procedures of the DEFM training process are provided in Supplementary Information Section 1.3.

## 3. Results

To demonstrate the performance of the proposed methods, the DEFM is first applied to three spatiotemporally chaotic systems: a 90-dimensional (90D) coupled Lorenz system under different parametric and noise conditions, the Lorenz 96 system, and the Kuramoto-Sivashinsky (KS) equation with inhomogeneity. In addition, we provide the Lyapunov times of these three systems in Section 2.2 of the Supplementary Information. Furthermore, the DEFM is tested on a series of real-world systems, including the cardiovascular inpatient dataset[29,30] (Section 2.4.5 in the Supplementary Information), the wind speed and solar irradiance dataset[31](Sections 2.4.4 and 2.4.6 in the Supplementary Information), the route of a typhoon center dataset[32] (Section 2.4.7 in the Supplementary Information), the traffic dataset[33] (Section 2.4.8 in the Supplementary Information), and the gene expression dataset[20] (Section 2.4.9 in the Supplementary Information). All these results show that the DEFM achieves better multistep-ahead prediction performance than the other five existing methods in terms of both accuracy and robustness. Therefore, considering its applicability to high-dimensional time series and reliable performance, the DEFM offers a new way to make accurate multistep-ahead forecasts for dynamic systems and has potential applications in many fields of artificial intelligence.

### 3.1 Prediction results obtained on a representative Lorenz dataset

To illustrate the mechanism and performance of the DEFM framework, we employ a 90D coupled Lorenz system[34] as the first representative model for generating synthetic time series data. The Lorenz system is formally expressed as:

$$\dot{Z}(t) = L(Z(t); P), \tag{12}$$

where the variables $Z(t) = (z_1^t, \dots, z_{90}^t)^T$, $L(\cdot)$ is a 90D function set of the Lorenz system, and the vector $P$ represents the time-invariant parameters. Specifically, the 90D coupled Lorenz system (Eq. (12)) is composed of 30 subsystems, and each subsystem is formed as:

$$\begin{cases} \dot{z}_{i,1} = \sigma^t \left( z_{i,2} - z_{i,1} \right) + c z_{i-1,i} \\ \dot{z}_{i,2} = \rho z_{i,1} - z_{i,2} - z_{i,1} z_{i,3} \\ \dot{z}_{i,3} = z_{i,1} z_{i,2} - \beta z_{i,3} \end{cases}, \tag{13}$$

where $i = 1, 2, \dots, 30$ denotes the $i$-th subsystem, $c z_{i-1,i}$ is the coupled factor for coupling the $i$-th subsystem with the previous subsystem and we set $i - 1 = 30$ for the situation $i = 1$ to keep the system closed, $c = 0.1, \rho = 28, \beta = \frac{8}{3}$ are constants of each subsystem. $\sigma^t$ is the factor determining whether the system is time-varying or time-invariant. A detailed description of the 90D coupled Lorenz system (Eq. (12)) is provided in Supplementary Section 2.1.

First, the DEFM is applied to a noise-free system (Eq. (12)) to predict the future dynamics of three randomly selected target variables $z_{k_1}, z_{k_2}$ and $z_{k_3}$. The prediction process is carried out with $n = 90$ dimensions, $m = 45$ as the length of the known series, and $S - 1 = 18$ as the prediction length. From a three-dimensional perspective (Figs. 2 (a)-(c)), the DEFM predicts the future trends (the red curves) of all the targets $\{z_{k_1}, z_{k_2}, z_{k_3}\}$. The DEFM produces multistep-ahead predictions not only for the single-wing case, i.e., the case in which the known/observed and unknown/to-be-predicted time series are distributed in the same wing of the attractor (as in Fig. 2(c)), but also for the cross-wing cases, i.e., the case in which the known/observed and unknown/to-be-predicted time series are distributed in two different wings of the attractor (as in Fig. 2 (a) and (b)). Figs. S1 (d)-(l) present the accurate single-target variable predictions for the three cases, i.e., (d)-(f) for $z_{k_1}$, (g)-(i) for $z_{k_2}$, and (j)-(l) for $z_{k_3}$. For all the target predictions $\{z_{k_1}, z_{k_2}, z_{k_3}\}$, the Pearson correlation coefficients (PCCs) between the eighteen predicted points and the real values are all above 0.999, while the root mean square errors (RMSEs) are all below 0.06, which demonstrates the high prediction accuracy of the DEFM in both the single-wing and the cross-wing cases. More prediction results and details of the synthetic Lorenz dataset are shown in Supplementary Section 2.4.1.

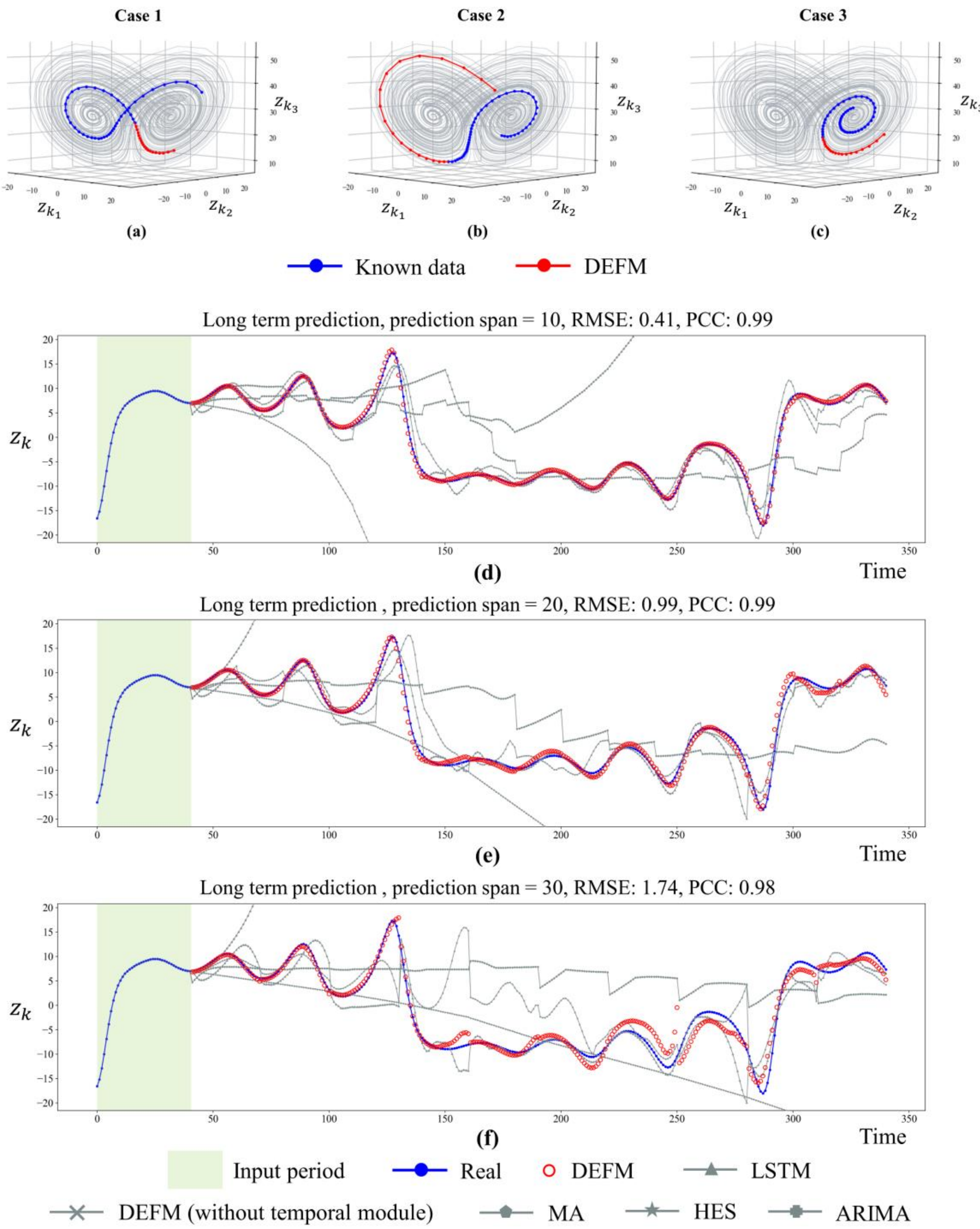

**Fig. 2.** The prediction performance achieved by the DEFM on a 90D synthetic Lorenz system Eq. (12). **(a)**-**(c)** The prediction results obtained for three randomly selected target variables $z_{k_1}$, $z_{k_2}$ and $z_{k_3}$ in 3D systems. The long-term prediction results yielded by the DEFM on the Lorenz system with different prediction spans are as follows: **(d)**: $S-1=10$, **(e)**: $S-1=20$, and **(f)**: $S-1=30$.

**Long-term prediction**

Due to its ability to obtain $(S-1)$-step-ahead future states in one prediction, the DEFM can be applied to make a long-term prediction for an $n$-dimensional system with an iterative scheme. Specifically, to forecast the long-term future states of a target variable $z_k$, e.g., $r(S-1)$-step-ahead future values $(r \geq 2)$, the trained DEFM is applied $r$ times with the updated inputs. For each $(S-1)$-step-ahead prediction, the input matrix $[Z]_{n\times m}$ is updated by two groups of data: one part contains the predicted series of a group of selected target variables, and the other part includes the observed time series of the remaining variables. As shown in Fig. 2 (d)-(f), the DEFM is applied to a 90D coupled Lorenz system (Eq. (12)), and it makes long-term predictions (300 steps) with different prediction spans $(S-1)$. It is seen that the DEFM still accurately forecasts the long-term future values when the span $S-1=10$ (Fig. 2 (d)), while the predictions deviate from the real values when the span grows to $S-1=20$ (Fig. 2 (e)) and $S-1=30$ (Fig. 2 (f)).

**Comparisons and robustness analysis**

The spatiotemporal architecture of the DEFM enables the full exploration of the temporally associated and spatially intertwined information contained in the observed time series of a high-dimensional system. To illustrate the superior performance of the proposed prediction method, the performance of the DEFM is compared with that of four traditional prediction methods, including the moving average (MA), Holt's exponential smoothing (HES)[8,9], the ARIMA[10], and LSTM[16]. Moreover, to illustrate the benefit brought by the temporal module, a simplified DEFM, which has the same structure as the full DEFM except for the temporal module, is also applied in the comparison. The details of these five methods are listed in Supplementary Section 2.3.

First, the DEFM and the other five prediction methods are applied to the dataset generated from the 90D coupled Lorenz system (Eq. (12)) with the length of unknown/to-be-predicted series in one prediction fixed to 18, i.e., $S-1=18$. Specifically, when the length of the known time series $m=80$, Fig. 3 (a) illustrates the prediction performances of all methods. It is seen that the DEFM accurately predicts the unknown values of the target variable (RMSE=0.058) and performs better than the other methods (the smallest RMSE=1.10) (Table S2). As the known time series becomes shorter, the DEFM still achieves high accuracy when $m = 60$ (RMSE=0.068, Fig. 3 (b)) and $m = 40$ (RMSE=0.117, Fig. 3 (c)), performing better than the other methods (the smallest RMSE=0.75 for the $m = 60$ cases, and the smallest RMSE=1.11 for the $m = 40$ cases (Table S2).

Second, the robustness of the DEFM is shown in an application involving the following 90D coupled time-varying Lorenz system:

$$\dot{Z}(t) = L\big(Z(t); P(t)\big). \tag{14}$$

where $L(\cdot)$ is the same 90D nonlinear function set of the Lorenz system in Eq. (12), but $P(t)$ is a time-varying/time-switching parameter vector, which changes when the time variable $t$ moves forward every 100 units. Notably, affected by external disturbances, most real-world systems are indeed time-varying rather than having constant parameters. Thus, the practical applicability of a prediction method should also be validated by its performance in such time-varying systems. As shown in Figs. 3 (d)-(f), when the length of the known time series decreases, i.e., (d) $m = 80$, (e) $m = 60$, and (f) $m = 40$, the DEFM performs stably and achieves the best accuracy among all methods in terms of both the RMSE (all below 0.076) and PCC (all above 0.999), while the smallest RMSE is 0.40 and the largest PCC is 0.996 for the other methods.

Moreover, for each known length value $m$ ranging from 40 to 80, comparisons are carried out in all cases involving the high-dimensional Lorenz system. As shown in Figs. 3 (g) and (h), the DEFM achieves the best prediction results with the highest PCC and lowest RMSE in terms of average performance. Notably, even with the short-term information observed from 40 time points, the DEFM still reaches a mean PCC of approximately 0.92 overall for randomly selected cases, while the mean PCCs and RMSEs of other methods are much worse, demonstrating the excellent ability of the DEFM to conduct short-term high-dimensional time series analyses. Compared with the stable performance of the DEFM, it is also worth noting that when the length of known series grows, the performance of LSTM becomes considerably poorer due to its limited memory[35]. Simultaneously, when examining DEFM under conditions of a fixed known length ($m = 80$) with variations in the multi-step prediction length (ranging from 20 to 40), the observed minimal fluctuations (Fig. S3) in prediction performance demonstrate the robustness of DEFM.

To further validate the robustness of the DEFM, a limited number of training cases and noisy conditions are applied. With different subsets of all 1000 cases selected from the 90D coupled Lorenz system (Eq. (12)), the DEFM is trained and tested. As shown in Fig. 3 (i), the DEFM still works well after being trained on only 20% of the selected cases, which suggests that the proposed framework can eliminate dependencies on a large amount of training data. In addition, noise is inevitable in real-world systems and disturbs the data analysis process. Trained by disturbed data generated from the 90D Lorenz system with white noise, the DEFM is employed to predict future values that are also perturbed by the noise. Fig. 3 (j) shows that the performance of the DEFM decreases gradually as the noise strength increases. However, even when the data are perturbed by considerably strong noise, i.e., when the variation of noise is approximately 2.0, the average PCC and RMSE are acceptable. We also investigated the sensitivity of the DEFM forecasting performance to different initialization methods and found that the DEFM performs well with several prevalent initialization methods[36,37], as detailed in Section 2.6 of the Supplementary Information.

From the above discussion, the robustness of the DEFM is demonstrated by its superior performance in both time-invariant and time-varying systems with even short-term known series. In other words, the DEFM is widely applicable to various high-dimensional systems and in different sampling conditions.

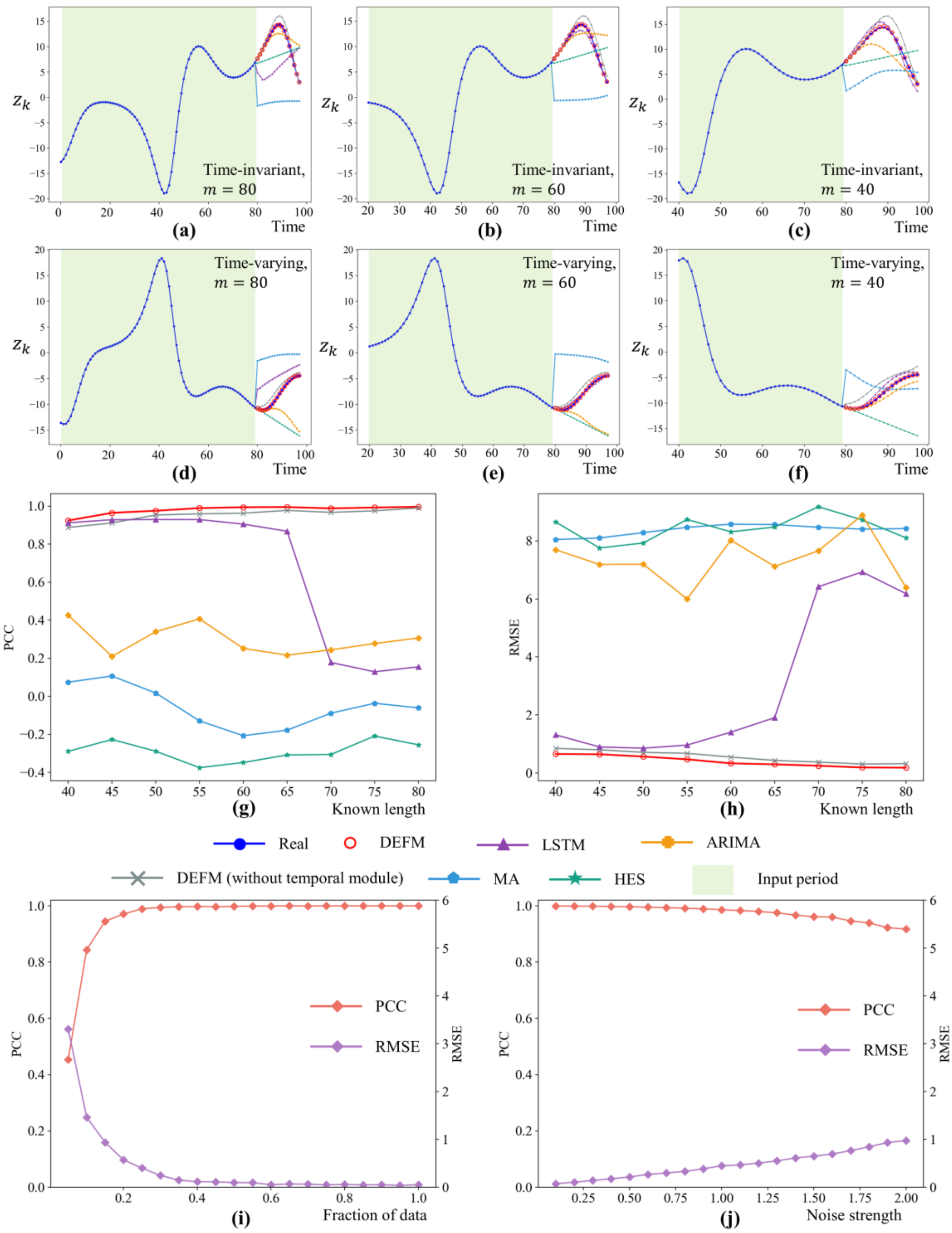

**Fig. 3.** Comparison and robustness analysis results were obtained with a fixed predicted time series length of $S-1=18$. Applying the six methods to the 90D coupled time-invariant Lorenz system (Eq. (12)), prediction results are obtained with different known lengths: **(a)** $m$=80, **(b)** $m$=60 and **(c)** $m$=40. The prediction results obtained for the 90D coupled time-varying Lorenz system (Eq. (14)) are as follows: **(d)** $m$=80, **(e)** $m$=60 and **(f)** $m$=40. For all 200 randomly chosen cases involving the Lorenz system, **(g)** the average Pearson correlation coefficient (PCC) and **(h)** the mean RMSE between the predictions and ground truths are determined as the length of the known time series increases, changing from 40 to 80. **(i)** The average PCC and RMSE between the DEFM predictions and the ground truths for cases with different data fractions. **(j)** The average PCC and RMSE between the DEFM predictions and the ground

truths for cases with different noise strengths.

### 3.2 Demonstration on the Lorenz 96 system

To demonstrate the effectiveness of the DEFM, we conduct a supplementary analysis using two additional representative numerical datasets, namely, the Lorenz 96 system[38] and the KS equation with inhomogeneity[17], and compare the DEFM with the other five methods.

The Lorenz 96 model, introduced by Lorenz in 1996[38], characterizes the macroscopic dynamics of the mid-latitude atmosphere. From a mathematical perspective, this model describes the temporal evolution of the components $x_j$, where $j \in \{0, 1, \ldots, J-1\}$, representing a spatially discretized atmospheric variable along a specific single latitude circle. The Lorenz 96 system can be formally defined as:

$$\frac{dx_j}{dt} = \left(x_{j+1} - x_{j-2}\right)x_{j-1} - x_j + F \tag{15}$$

for $j \in \{0, 1, \ldots, J-1\}$, where it is assumed that $x_{-1} = x_J, x_{-2} = x_{J-1}$, and $F$ is a forcing constant that controls the chaotic nature of the system. In this study, we set $J = n = 60, F = 5$ to generate the numerical dataset. The DEFM is employed to forecast the Lorenz 96 system with a known length of $m = 40$ and a prediction horizon of $S - 1 = 18$. As the examples shown in Figs. 4 (a)-(c), the DEFM outperforms the other five methods, achieving the highest prediction accuracy. The predicted results closely align with the actual values, exhibiting an impressive near-perfect level of alignment. Furthermore, we systematically evaluate the DEFM and the other compared methods, employing the average PCC and RMSE metrics calculated on the synthetic Lorenz 96 dataset. The results obtained for these two metrics across different methods are visually represented in Fig. 4 (g) and Fig. 4 (h), respectively. We can observe that the DEFM achieves the best performance in terms of both evaluation metrics. Notably, the DEFM exhibits a 67.5% PCC improvement over the second-best approach (ARIMA), which is accompanied by a significant RMSE error reduction of 75.6%.

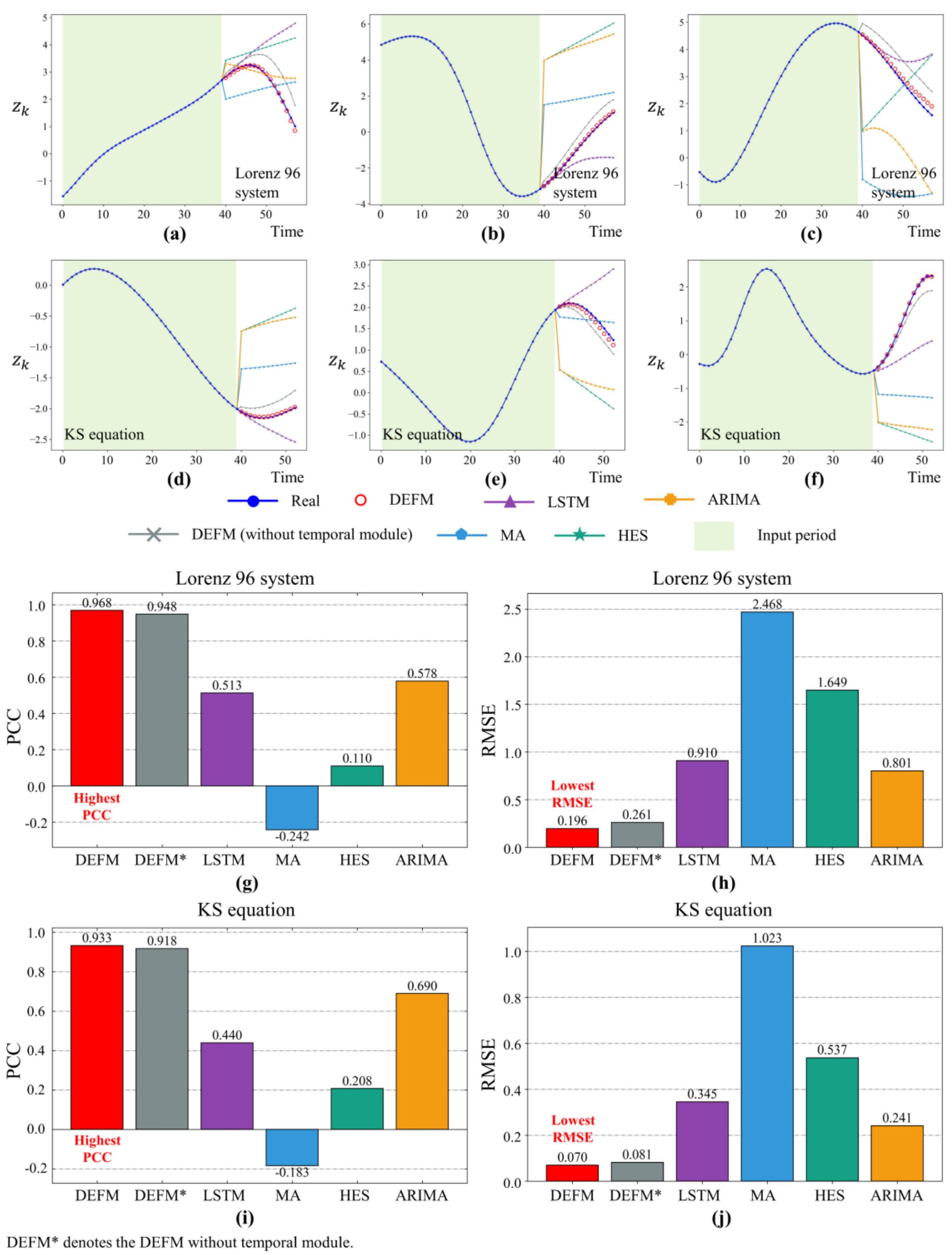

**Fig. 4.** Validation of the performance of the DEFM on the Lorenz 96 system and the KS equation. The forecasting results are obtained by six different models on the Lorenz 96 system under a known length of $m = 40$ and a prediction horizon of $S - 1 = 18$ **(a-c)** and on the KS equation under $m = 40$ and $S - 1 = 13$ **(d-f)**. Performance comparisons are conducted among the six models using the PCC and RMSE metrics attained on the Lorenz 96 system **(g, h)** and the KS equation **(i, j)**, respectively. Note that DEFM* represents the DEFM model without a temporal module.

In general, the dimensionality of a system significantly influences predictive performance. We conducted a thorough examination of the performance of DEFM on the Lorenz 96 system (Eq. (15)) under varied dimensional conditions by incrementing the value of $J$. Additionally, the influence of the forcing term $F$ has been meticulously considered, with specific settings of $F = 4$ and $F = 5$. Note that a higher value of $F$ indicates an augmented chaotic nature. For all the conditions, the known and prediction lengths of DEFM are fixed at $m = 40$ and $S - 1 = 18$, respectively. Table 1 summarizes the evaluation results of DEFM applied to the Lorenz 96 system, encapsulating the DEFM's performance across distinct forcing terms and increasing dimensions. When $F = 4$, the Lorenz 96 system exhibits weak chaotic behavior, and the performance degradation of DEFM with incremental system dimensions is relatively minor. However, in the case of $F = 5$, the predictive RMSE errors of DEFM tend to escalate with the increase of system dimensions, reflecting the heightened complexity of the system. Notably, as the system dimension rises, although there is an augmentation in spatial information for the target variable, a concomitant rise in the number of irrelevant variables incorporated by the model introduces a surplus of extraneous information and noise into the forecasting task.

**Table 1.** Performance of DEFM on Lorenz96 system under different parameter settings.

| Dimension $J$ | $F = 4$ | | $F = 5$ | |
|---|---|---|---|---|
| | RMSE | PCC | RMSE | PCC |
| 60 | 0.010 | 0.999 | 0.196 | 0.968 |
| 100 | 0.010 | 0.999 | 0.237 | 0.955 |
| 150 | 0.045 | 0.997 | 0.270 | 0.925 |
| 200 | 0.012 | 0.999 | 0.310 | 0.919 |
| 250 | 0.055 | 0.997 | 0.327 | 0.916 |
| 300 | 0.025 | 0.999 | 0.359 | 0.911 |

To explore the prediction horizon of DEFM on the Lorenz 96 model ($F = 5,\ J = 60$), we trained 60 target variable-specific models for each variable within the system. Fig. 5 presents the prediction horizon outcomes for a Lorenz 96 system ($F = 5, J = 60$), where Fig. 5 (a) presents the numerical solution of Eq. (15), Fig. 5 (b) depicts the predictions iteratively obtained from the ensemble of 60 models, and Fig. 5 (c) illustrates the prediction error (Fig. 5 (b) subtracted from Fig. 5 (a)). Evidently, the prediction error remains low for approximately 5 Lyapunov times, underscoring the effectiveness of DEFM in the context of the Lorenz 96 model. Furthermore, comprehensive analyses on the impact of the observability of the complete system state on the prediction performance of DEFM are presented in Section 2.4.3 of Supplementary Information.

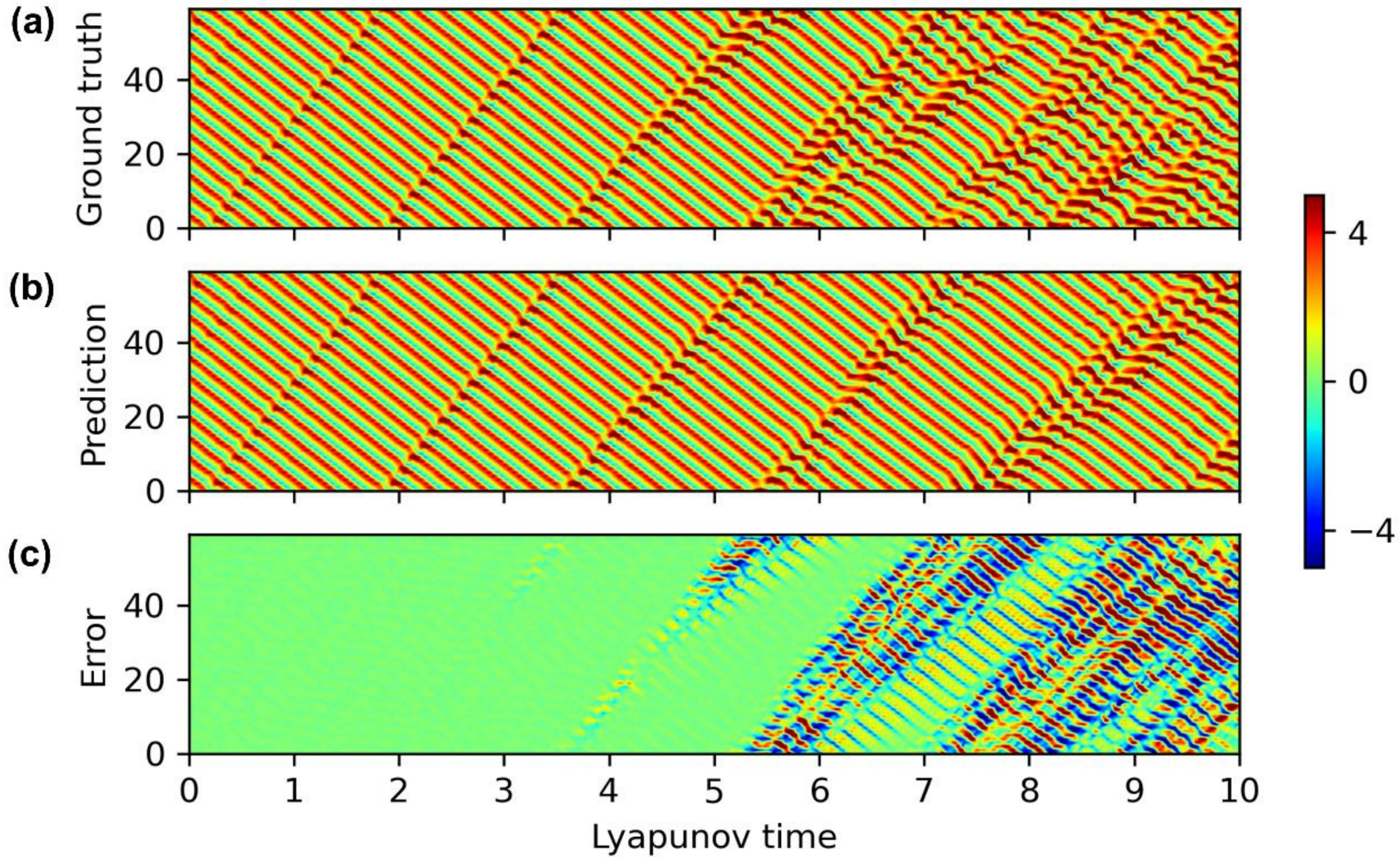

**Fig. 5.** Visualization of the prediction outcomes for the entire Lorenz 96 system ($F = 5,\ J = 60$). **(a)** Ground truth derived from the Lorenz 96 system. **(b)** Iteratively obtained predictions from DEFM. **(c)** The prediction error (derived by panel (a) minus panel (b)). The ordinate axis represents all the system variables, while the abscissa axis represents the time $t$ scaled in units of one Lyapunov time, as time $t$ is multiplied by the largest Lyapunov exponent of the system.

### 3.3 Validation on the Kuramoto-Sivashinsky equation with inhomogeneity

To examine another spatiotemporally chaotic system, we consider the modified KS equation, incorporating an additional spatial inhomogeneity term[17]:

$$y_t = -yy_x - y_{xx} - y_{xxxx} + \mu \cos\left(\frac{2\pi x}{\lambda}\right). \tag{16}$$

The scalar field $y(x,t)$ is periodic within the interval $[0, L)$, where $L$ is an integer multiple of $\lambda$. Note that the dimensionality of the attractor directly depends on the dimensionless parameter $L$, scaling linearly with $L$ for sufficiently large values of $L$. The KS equation (Eq. (16)) is integrated on a grid comprising $Q$ equally spaced points with a temporal increment of $\Delta t = 0.25$. This integration yields a simulated dataset containing $Q$ time series, thereby establishing a $Q$-dimensional ($n = Q$) system. Following the approach outlined in reference[17], the modified KS equation (Eq. (16)) used here adopts the following parameter values: $L = 200, Q = 512, \mu = 0.01,$ and $\lambda = 100$. In the context of this chaotic system, the forecasting results obtained from the DEFM reveal slight disparities when compared to the actual solutions (Figs. 4 (d)-(h)). In particular, akin to its performance on the Lorenz 96 system, the DEFM showcases significantly superior average PCC and RMSE values for its KS equation prediction results in comparison with the other methods (Fig. 4 (i) and Fig. 4 (j)). Additionally, the results and analyses for the prediction horizon on the modified KS systems with varying system sizes ($L = 100, 200,$ and $400$) are presented in Section 2.4.10 of the Supplementary Information.

Specifically, our analysis reveals that the DEFM predicts up to approximately 1 Lyapunov time, as shown in Fig. S14 of the Supplementary Information, which is less effective than the reservoir computing method proposed by Pathak et al. in reference[17]. In fact, the approach proposed in reference[17] is easier to implement and proves to be a better method when there is less training data available, for making longer-term predictions or when the training samples are from spatiotemporally chaotic systems like the KS equation or Lorenz 96 system. In terms of accuracy, the DEFM excels in short-term time-series forecasting. Meanwhile, in scenarios where variables exhibit more complex nonlinear relationships (such as with real-world data), the DEFM, based on the STI transformation, is more effective in predicting the future information for each target variable.

## 4. Conclusion

In this paper, we present a novel dynamics-based machine learning model, i.e., a DEFM,

to make multistep-ahead predictions of future information based on high-dimensional time series data. The key of the DEFM method is to represent the STI equation by solving a nonlinear function that maps from the known information (the high-dimensional time series of a complex system) to the future information (the time series of a target variable). Based on the delay embedding scheme, which enlarges the training sample size by including labeled embeddings and unlabeled future values (Fig. 1), the DEFM makes use of high-dimensional time series data with its three spatiotemporal modules. Trained in a self-supervised way, the DEFM fits the nonlinear function well and accurately predicts the future dynamics of the target variable in a multistep-ahead manner. Moreover, due to its ability to obtain multistep-ahead future states in one prediction, the DEFM can be applied to make long-term predictions with an iterative scheme. As a universal deep learning method, the DEFM is not only effective in forecasting in dynamical systems with explicit models, but also exhibits satisfactory performances on various real-world datasets, demonstrating its applicability in periodic-like datasets such as the single-wing cases of a Lorenz system (Case 3 in Fig. 2) and solar irradiance (Fig. S10), nonperiodic datasets such as the two-wing cases of a Lorenz system (Cases 1 and 2 in Fig. 2) and typhoon positioning (Fig. S11), and even in highly fluctuating cases such as those involving wind speed (Fig. S8) and traffic flow (Fig. S12). In addition, a comparison with previously developed methods[23,25,39] that incorporate the STI transformation is expounded upon in Section 2.7 of the Supplementary Information, highlighting the advancements achieved by the DEFM.

The DEFM achieves a prediction horizon of approximately 5 Lyapunov times for the Lorenz 96 model, in contrast to around 1 Lyapunov time for the KS system, as shown in Figs. 5 and S14, respectively. Upon further analysis, we hypothesize that the discrepancy in predictive performance between the KS case and the Lorenz 96 system may be attributed to the inherent characteristics of the data generated from each model. Specifically, the data generated from the KS model exhibits stronger temporal variability and periodicity compared to the time series from the Lorenz 96 system. As our DEFM method relies on recent short-term time series for prediction, the more pronounced variability in the KS system data may pose challenges for accurate forecasting. We will explore this hypothesis further in our future work and provide additional insights into this observation.

The DEFM method possesses obvious advantages. On the one hand, the well-designed architecture of the DEFM composed of temporal, spatial, and merge modules makes it capable of fully exploring both the spatial interactions and the temporally associated information among high-dimensional variables. On the other hand, compared with traditional prediction methods, the DEFM exhibits higher accuracy and robustness in time-varying and noisy systems (Fig. 3 and Fig. 4), which are prevalent in practical applications. In view of these benefits, the DEFM opens a new multistep-ahead prediction avenue to scenarios with high-dimensional time series data and is thus potentially useful in a wide variety of real-world systems.

## Supplementary Material

See the Supplementary Material for the details of Takens' delay embedding theory[21,22] and the training pipeline of the DEFM. Additionally, numerous details and analyses related to the experiments are provided in the Supplementary Material, such as, thorough descriptions for the validation datasets and methods used for comparison, supplementary results illustrating the superiority of DEFM, and a sensitivity analysis concerning initialization methods of DEFM.

## Acknowledgements

This work was supported by National Natural Science Foundation of China (Grant numbers T2341022, 12322119, 62172164, and 12271180), Guangdong Provincial Key Laboratory of Human Digital Twin (2022B1212010004).

## Author declarations

### Conflict of Interest

The authors have no conflicts to disclose.

**Author Contributions**

**Hao Peng:** Conceptualization (equal); Formal analysis (equal); Investigation (equal); Methodology (equal); Visualization (equal); Writing – original draft (equal); Writing - review & editing (equal). **Wei Wang:** Formal analysis (equal); Investigation (equal); Writing - review & editing (equal). **Pei Chen:** Conceptualization (equal); Formal analysis (equal); Investigation (equal); Methodology (equal). **Rui Liu:** Conceptualization (lead); Writing - review & editing (equal).

## Data availability

All data needed to evaluate the conclusions are present in the paper and/or the Supplementary Materials. All codes are available at https://github.com/Peng154/Delay-Embedding-based-Forecast-Machine.

# References

[1] D.J. Lockhart, and E.A. Winzeler, "Genomics, gene expression and DNA arrays," Nature **405**(6788), 827–836 (2000).

[2] M.M. Rienecker, M.J. Suarez, R. Gelaro, R. Todling, J. Bacmeister, E. Liu, M.G. Bosilovich, S.D. Schubert, L. Takacs, G.-K. Kim, and others, "MERRA: NASA's modern-era retrospective analysis for research and applications," J. Clim. **24**(14), 3624–3648 (2011).

[3] J. Fan, F. Han, and H. Liu, "Challenges of big data analysis," Natl. Sci. Rev. **1**(2), 293–314 (2014).

[4] H. De Jong, "Modeling and simulation of genetic regulatory systems: a literature review," J. Comput. Biol. **9**(1), 67–103 (2002).

[5] R.R. Stein, V. Bucci, N.C. Toussaint, C.G. Buffie, G. Rätsch, E.G. Pamer, C. Sander, and J.B. Xavier, "Ecological modeling from time-series inference: insight into dynamics and stability of intestinal microbiota," PLoS Comput. Biol. **9**(12), (2013).

[6] I.H. Stevenson, and K.P. Kording, "How advances in neural recording affect data analysis," Nat. Neurosci. **14**(2), 139 (2011).

[7] M.X. Cohen, *Analyzing Neural Time Series Data: Theory and Practice* (MIT press, 2014).

[8] C.C. Holt, "Forecasting seasonals and trends by exponentially weighted moving averages," Int. J. Forecast. **20**(1), 5–10 (2004).

[9] R.G. Brown, in *Oper. Res.* (INST OPERATIONS RESEARCH MANAGEMENT SCIENCES 901 ELKRIDGE LANDING RD, STE 400, LINTHICUM HTS, MD

21090-2909, 1957), pp. 145–145.
[10] G.E. Box, and D.A. Pierce, "Distribution of residual autocorrelations in autoregressive-integrated moving average time series models," J. Am. Stat. Assoc. **65**(332), 1509–1526 (1970).
[11] A.G. Parlos, O.T. Rais, and A.F. Atiya, "Multi-step-ahead prediction using dynamic recurrent neural networks," Neural Netw. **13**(7), 765–786 (2000).
[12] C.L. Giles, S. Lawrence, and A.C. Tsoi, "Noisy time series prediction using recurrent neural networks and grammatical inference," Mach. Learn. **44**(1–2), 161–183 (2001).
[13] A. Lazar, G. Pipa, and J. Triesch, "Fading memory and time series prediction in recurrent networks with different forms of plasticity," Neural Netw. **20**(3), 312–322 (2007).
[14] J.T. Connor, R.D. Martin, and L.E. Atlas, "Recurrent neural networks and robust time series prediction," IEEE Trans. Neural Netw. **5**(2), 240–254 (1994).
[15] Z. Karevan, and J.A. Suykens, "Transductive LSTM for time-series prediction: An application to weather forecasting," Neural Netw., (2020).
[16] S. Hochreiter, and J. Schmidhuber, "Long short-term memory," Neural Comput. **9**(8), 1735–1780 (1997).
[17] J. Pathak, B. Hunt, M. Girvan, Z. Lu, and E. Ott, "Model-free prediction of large spatiotemporally chaotic systems from data: A reservoir computing approach," Phys. Rev. Lett. **120**(2), 024102 (2018).
[18] J. Shotton, A. Fitzgibbon, M. Cook, T. Sharp, M. Finocchio, R. Moore, A. Kipman, and A. Blake, in *CVPR 2011* (2011), pp. 1297–1304.
[19] B. Johnson, M. Gomez, and S.B. Munch, "Leveraging spatial information to forecast nonlinear ecological dynamics," Methods Ecol. Evol. **12**(2), 266–279 (2021).
[20] Y. Wang, X.-S. Zhang, and L. Chen, "A network biology study on circadian rhythm by integrating various omics data," OMICS J. Integr. Biol. **13**(4), 313–324 (2009).
[21] T. Sauer, J.A. Yorke, and M. Casdagli, "Embedology," J. Stat. Phys. **65**(3–4), 579–616 (1991).
[22] F. Takens, in *Dyn. Syst. Turbul. Warwick 1980* (Springer, 1981), pp. 366–381.
[23] H. Ma, S. Leng, K. Aihara, W. Lin, and L. Chen, "Randomly distributed embedding making short-term high-dimensional data predictable," Proc. Natl. Acad. Sci. **115**(43), E9994–E10002 (2018).
[24] H. Ma, T. Zhou, K. Aihara, and L. Chen, "Predicting time series from short-term high-dimensional data," Int. J. Bifurc. Chaos **24**(12), 1430033 (2014).
[25] P. Chen, R. Liu, K. Aihara, and L. Chen, "Autoreservoir computing for multistep ahead prediction based on the spatiotemporal information transformation," Nat. Commun. **11**(1), 4568 (2020).
[26] A. Vaswani, N. Shazeer, N. Parmar, J. Uszkoreit, L. Jones, A.N. Gomez, \Lukasz Kaiser, and I. Polosukhin, in *Adv. Neural Inf. Process. Syst.* (2017), pp. 5998–6008.
[27] Q. Chen, H. Zhao, W. Li, P. Huang, and W. Ou, in *Proc. 1st Int. Workshop Deep Learn. Pract. High-Dimens. Sparse Data* (2019), pp. 1–4.
[28] S. Li, X. Jin, Y. Xuan, X. Zhou, W. Chen, Y.-X. Wang, and X. Yan, in *Adv. Neural Inf. Process. Syst.* (2019), pp. 5244–5254.
[29] T.W. Wong, T.S. Lau, T.S. Yu, A. Neller, S.L. Wong, W. Tam, and S.W. Pang, "Air

pollution and hospital admissions for respiratory and cardiovascular diseases in Hong Kong.," Occup. Environ. Med. **56**(10), 679–683 (1999).
[30] J. Fan, W. Zhang, and others, "Statistical estimation in varying coefficient models," Ann. Stat. **27**(5), 1491–1518 (1999).
[31] Y. Hirata, and K. Aihara, "Predicting ramps by integrating different sorts of information," Eur. Phys. J. Spec. Top. **225**(3), 513–525 (2016).
[32] Informatics, N. I. o., "Digital Typhoon: Cyclone 201820 (Marcus) - General Information (Pressure and Track Charts)," (n.d.).
[33] Y. Li, R. Yu, C. Shahabi, and Y. Liu, "Diffusion convolutional recurrent neural network: Data-driven traffic forecasting," ArXiv Prepr. ArXiv170701926, (2017).
[34] J.H. Curry, "A generalized Lorenz system," Commun. Math. Phys. **60**(3), 193–204 (1978).
[35] I. Sutskever, O. Vinyals, and Q.V. Le, in *Adv. Neural Inf. Process. Syst.* (2014), pp. 3104–3112.
[36] X. Glorot, and Y. Bengio, in *Proc. Thirteen. Int. Conf. Artif. Intell. Stat.* (2010), pp. 249–256.
[37] K. He, X. Zhang, S. Ren, and J. Sun, in *Proc. IEEE Int. Conf. Comput. Vis.* (2015), pp. 1026–1034.
[38] E.N. Lorenz, in *Proc Semin. Predict.* (Reading, 1996).
[39] H. Peng, P. Chen, R. Liu, and L. Chen, "Spatiotemporal information conversion machine for time-series forecasting," Fundam. Res., (2022).